Institute for Nuclear Research and Nuclear Energy
Bulgarian Academy of Sciences

# Seminar Proceedings

# Possible correlation between electromagnetic earth fields and future earthquakes

Organizing committee:
S.Cht.Mavrodiev, C. Thanassoulas

23-27 July, 2001
Sofia

# Possible correlation between electromagnetic earth fields and future earthquakes

The support of Institute Administration for the Seminar and this Proceeding is acknowledged. The attempt of Mr. Maragos from European Commission and some persons from the Bulgarian Government Defense commission against natural and antropogenic incidents to sponsor us was very inspiriting.

Prepared by S.Cht.Mavrodiev
Institute for Nuclear Research and Nuclear Energy
Bulgarian Academy of Sciences
72 Boul. Tzarigradsko shaussee
1784 Sofia, Bulgaria

Director Prof. Dr. J. Stamenov
Tel: +359 2 75 80 32
        359 2 7144 647
Fax:   359 2 75 50 19
Telex:  24 368

Institute for Nuclear Research and Nuclear Energy
Bulgarian Academy of Science
Department "Astrophysical Objects and Environment"

**Announcement- Invitation**

Dear Researchers,

The Department "Astrophysical Objects and Environment" intends to organize a 3-day seminar at Sofia, 23-26, July 2001, under the title:

**"The possible correlation between electromagnetic earth surface fields and future local earthquakes".**

In the seminar the work done by the Balkan researchers and the ideas for future investigations of the topic, will be presented and discussed aiming as follows:

**A.** Establish a regional Balkan monitoring network of the earth's electrical field targeting to earthquake prediction by utilizing the method presented in web-sites: http://users.otenet.gr/~thandin/index.htm
http://www.om2.inrne.bas.bg/geomagflde.htm.

**B**. Formulate a research project on this topic and discuss the details, possible fund resources and participants.

**C**. NGO "Balkans, Harmonic Existence and Science"

The general opinion is that our countries face a rather increased seismic activity lately that justifies fostering the research and possible applications on this topic.
At this stage of planning it is made clear that no funds are available at all and the seminar participants are kindly asked to cover their own expenses by their own means. The hotel accommodation will be in the framework of 20 US$/day with breakfast, the lunch and dinner 10 US$/day.
We are looking for sponsors.
At the present time we would like to know your point of view and your possible participation. If You will come, please mail the text of Your talk for the Seminar proceedings.
As an example of the electrical field method is the case of Hios, Greece EQ (5.6R 10/06/01). Dr. Thanassoulas, C. : thandin@otenet.gr.

S. Cht. Mavrodiev

Director        Prof. Dr. J. Stamenov,
INRNE, Bulg. Ac. of Sc., Sofia, Bulgaria

Organizing Committee:
                Ass. Prof., Dr. S.Cht.Mavrodiev, INRNE, BAS, Bulgaria
                Dr. C. Thanassoulas, Greece

**Program:**

**Monday, 23 July**

| | | | | |
|---|---|---|---|---|
| Arriving and accommodation | | 12:00 | | |
| Lunch | | 12:30 | - | 13:30 |

Opening

| | | | | |
|---|---|---|---|---|
| Prof.Dr.J.Stamenov | Opening adress | 14:00 | - | 14:15 |
| S.Cht.Mavrodiev | About the Seminar | 14:15 | | 14:30 |

A.  Dr. C. Thanassoulas,  Earthquake prediction based on electrical signals
recorded on ground surface                15:00  -      18:00
Dinner                                    20:00

**Tuesday, 24 July**

A.  Discussion
B.Rangelov    EARTHQUAKE ENVIRONMENT, MONITORING AND
PREDICTION EXPERIENCE IN BULGARIA      09:00  -      10:00

B. Ranguelov, A. Kies      Rn measurements and geodynamics in/around
the seismic station Kroupnik (SW Bulgaria)      10:00      11:00

Б.Рангелов The today science and earthquakes predeiction (in Bulgarian)
                                          10:00      11:00
S.Cht.Mavrodiev            The possible correlation's between electroma-
gnetic earth surface fields and future local earthquakes
                                          11:00  -      12:00
Lunch                                    12:30  -      13:30
A.  Conclusion resume                    15:00  -      18:00
B.  Discussion                           18:00  -      19:00
Dinner                                    20:00

**Wednesday, 25 July**

B.  Discussion                           08:00  -      09:00
Preliminary Project Version              09:00  -      12:00
Lunch                                    12:30  -      13:30

C.  NGO  "Balkans, Harmonic Existence and Science"  14:00  -      15:00
Closing ceremony                         15:00  -      16:00

mavrodi@inrne.bas.bg

# EARTHQUAKE ENVIRONMENT, MONITORING AND PREDICTION EXPERIENCE IN BULGARIA

**Boyko Ranguelov,** Geophysical Institute - BAS

## Earthquake occurrence

The territory of Bulgaria is under regional general extension regime due to the collision between African and European plates. Most strong earthquakes have a normal faulting observed as well on the surface ruptures during the strong earthquake occurrence of the Earth's surface. Some of them are combined with a strike-slip movements, especially for the shallower and not very strong seismic events. The catalogue has been completed since I-st century BC. The catalogue is incomplete and consists of historical part (I-st BC - beginning of the 20-th century), macroseismic part (end of 19-the century - almost present days) and instrumental part (beginning of the 20-th century to the present days). The idea for precursory monitoring test-site development started in Bulgaria during early 80-ies, but up to now, due to the lack of money, only some experimental methodology measurements have been executed, including the measurements of the earth electric potentials, Rn measurements in the soil, air and waters near Krupnik seismic stations and some other geodynamic investigations connected by the extensometer's measurements, GPS geodetic monitoring and some underground hydrology research.

The most significant seismic events during the well documented times are as follows:

13[th] September 1856 - Sofia region, estimated magnitude over 6.5, observed intensity - IX MSK, coseismic normal fault, boiled sands, mineral spring appeared, large destruction of the buildings, some deaths and many injured reported, long lasted aftershock activity - more than 5 months.

31[st] March, 1901 - Shabla-Kaliakra region, epicenter in the aquatory of the Black sea (NE coast), estimated magnitude - 7.1, observed intensity - up to X MSK, a foreshock (M~4) reported several ours prior the main shock, many coseismic and post-seismic events reported (landslides, stonefalls, liquefaction, tsunami effects - up to 3 meters, about 5 years aftershock activity). Many destructed houses, deaths and injured reported.

4[th] April 1904 - Kresna-Kroupnik region, two very strong shocks in time domain of about 20 minutes occurred (M=7.2 and M=7.8). Intensities up to X MSK reported. All coseismic and post-seismic events observed and reported - landslides, stonefalls, surface normal faulting (the river has been barraged and a lake observed), liquefaction, spring appearance, etc. Many deaths and injured reported. Big destruction of the houses. Long lasted aftershock sequence - more than 7 years. This is the most active part of Bulgaria up to now.

16[th] June 1913 - Gorna Oriahovitza region. Magnitude 7.0, intensity - up to IX degree MSK. Large destruction, many deaths and injured reported. Landslides, sand boils, liquefaction and aftershocks reported. 7[th] December 1986 a magnitude 5.7 earthquake occurred at the same region. Large destruction and a few deaths and injured have been reported.

14[th] and 18[th] April 1928 - again two shocks (M=6.8 and 7.0) with IX and X degree MSK reported. Many liquefactions, sand boils, landslides, surface normal faulting reported and geodetically measured (mention in the Richter's "Elementary seismology"). More than 120 deaths and several hundreds injured. More than three years aftershock's activity.

During the last years the Vrancea 1977 (M7.2), Velingrad 1977 (M5.3) and Strazhica 1986 (M 5.7) earthquakes occurred generating light, moderate and huge destruction in the respective macroseismic areas and some of them accompanied by victims (especially the Vrancea shock) .

**Seismic zonation**
According of the above described seismic manifestations as well as the tectonic, geological, geophysical and geodetic information at the end of 70-ties and the beginning of 80-ies the seismic zoning maps have been created. The general output are two sets of maps - the expected maximum magnitudes zoning map and the series of shakeability maps for the different time periods. The one of 1000 years has been accepted as a basic map for the seismic rules and code in Bulgaria. The maximum expected magnitude for the territory of Bulgaria reach up to 8.0 in the region of Kresna-Kroupnik (SW Bulgaria) and around the Kaliakra zone (NE Bulgaria). There are no almost aseismic zones on the territory. To the intensities on the 1000 years shakeability map the design coefficients have been attached. Now the seismic code of Bulgaria is modified according EUROCODE 8. There are a lot of discrepancies in the border regions of Bulgaria and other countries about the expected magnitudes and seismic intensities. Thus it is clear that the seismic zoning must be improved. This is due as well to the new and more reliable information obtained during the last years of investigations.

**Seismic monitoring**
The first modern seismic station has been established in Sofia in 1905 (three component Ommory-Bosch instrument) immediately after the strong events (M=7.2 and 7.8) in April 1904. There is a clear line that the development of the instrumentation follows the occurrence of the strong events on the territory of Bulgaria. After the strong events in April 1928 (M=6.7 and 7.0) the leveling measurements have been executed and surface dislocations very well documented (cited by Ch. F. Richter in his book "Elementary Seismology") the Wiehert two horizontal component mechanical seismograph (1000 kg) has been installed (1905). The service has been located during the years in the different institutions, but always governmental. The further development depends on the introduction of a Russian equipment (after the Second World War) - mainly electromagnetic instruments - SKD (Kirnos mid-60-ties) and SKM (short period instruments). The two component mechanical seismograph Krumbach has been installed in the early 60-ties - in Dimitrovgrad station. At the beginning of 80-ties the recent network have been established (S-13 Teledyne short period velocitygraphs - S13). Now in operation we have 13 regional analogue stations, 2 local networks with 3 and 4 stations respectively (one near the NPP and one near the big salt body under exploitation, which is supplied as well with 4 autonom three component digital accelerographs produced by GeoSIG, one broadband MEDNET station and one more network near the future second NPP site (under construction). All stations works in the near real time, by a telephone connections mainly. The hypocentres determinations are made by the HYPO73 software. The local magnitude duration determination have been made by a normalization to Ms. There is no earthquake mechanisms solutions, moment tensor magnitude or dynamic properties estimations in the routine practice of the network. The obtained data have been stored on the smocked paper (for the old instruments), pen ink records for the network registration and photo paper for the Russian equipment. Up to now there are not available digital records except the S13 - PDP old computer with 50 Hz digitization. The strong motion network also exists (equipped with SMA1, and other analogue or digital instruments) to support the practical application in the construction of buildings and facilities. It is managed by the CLSMSEE-BAS. Some of the produced data has been published several years ago. The present condition of the network is not very good, because of the lack of funding for the maintaining actions. Very often the advice of the scientists are neglected by the governmental institutions and their opinion is not taken into consideration before the strong earthquake occurrence. After that usually is too late. The nearest task is to transform all analogue recording stations into digital ones.

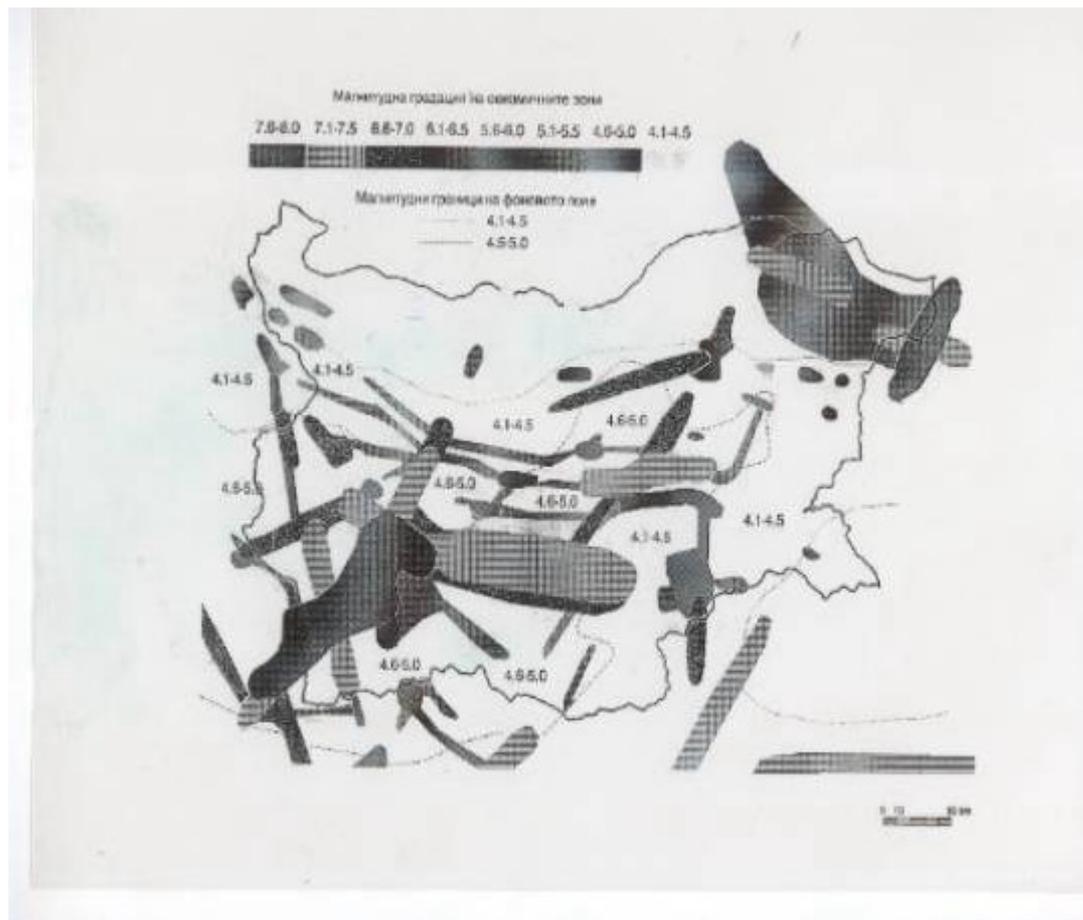

**Other monitoring data and correlation with earthquakes**
During the mid 80-ties a special stochastic model for earthquake prediction has been developed in Bulgaria. The application of this model for the expected future earthquakes gives some reasonable results in the time domain of 5-7 years with a successful prediction of about 60% efficiency. The model have been applied for some prediction purposes about the Mexican earthquakes. Some success also has been mentioned, but in the mid-term prediction. The next step of the Bulgarian earthquake prediction experience was connected with the establishment of the He concentration measurements in the borehole near Simitly. Here no precursory phenomena have been observed but some coseismic correlation established. They have been supported as well by the extensometer measurement in the same area. Following the world experience in the end of the 80-ties the earth electric field measurements started in Vitosha Geophysical observatory. Due to the lack of money they have been canceled. Now the reestablishment of these measurements started in the Geophysical Institute, together with the Institute of Space research. The last established measurements are connected with the Rn monitoring in the Kroupnik seismic station. No positive correlation with the earthquakes is observed.

**Recommendations**
It is important to recommend wide co-operation among the scientists of the Balkan Countries region for the seismological, building's structural, prevention and prediction research. The area is under strong influence of the earthquakes and other geodynamic disasters. The improvement of the instrumental networks connected by the modern communication networks are necessary conditions for the successful development of the region and the seismic protection and prediction measurements. We consider that the aims of the seismology, earthquake engineering and the people protection and prevention is a governmental task and the countries must support them by funding from different sources.

**Rn measurements and geodynamics in/around the seismic station Kroupnik (SW Bulgaria)**

Boyko Ranguelov - Geophysical Institute, BAS, Sofia
Antoine Kies - Centre Universitaire, Luxembourg

**Introduction**
During the last several years in Kroupnik seismic station located in the most seismic active area in Southwest Bulgaria, continuous radon measurements have been performed. The aim of these measurements is to use radon as a tracer gas for general geodynamical purposes. The radioactive noble gas radon Rn-222 has been monitored in air, in soil and in some water samples.

The results obtained so far have been reported during several meetings in Bulgaria and abroad.

**Devices for Rn monitoring**
For the measurements of ambient airborne radon concentrations the passive radon monitor ALPHAGUARD is used. This device is able to measure continuously, besides radon concentrations also the air pressure and the air temperature at the measurement place. Additionally an external thermometer, deposited in a proper place records the outside temperature. The time sampling interval is one measurement each hour (sometimes each 10 minutes).

In a next step the ALPHAGUARD device was located in a more confined place at the bottom of the instrument room situated in the basement of the station.
In order to minimise the influence of outside temperature, a one meter borehole was drilled at the bottom of the instrument room. Unfortunately, water seeping into the borehole prevented measurements in the borehole so far.

Radon in soil air has been analysed at different locations around the geophysical station. Soil-air is sucked out of tubes, hammered in the soil to a depth of 1 meter. The radon concentration is measured in scintillation cells (Lucas cells), by the photomultiplayer.

Water samples have been taken from the sources located in Simitly and the surroundings of the Simitly Graben. In Simitly the radon concentration of hot waters flowing out of two boreholes was measured. One borehole is situated close to the bath; the other is recently drilled, located nearby across the railway in a closed site where the waters here are used for the vegetable growing (orangeries). The water samples were analysed by liquid scintillation, performed in a Sofia laboratory and in Luxembourg.

Some experimental measurements have been made in Breznica village on an artesian mineral water source. High humidity and a water temperature of 60 °C prevented the use of a classical continuously working radon-in-water monitor. It is planned to install at this location an appropriate device able to work under those hard conditions.

**Geodynamic investigations**

At the beginning the continuous complex measurements have been established in the basement of the Kroupnik seismic station [2]. The registered concentrations show high peaks in radon content, up to 50 kBq/m3 in the air of the instrument room - [3]. The initial purpose of the measurements is the study of a possible influence of local earthquakes to radon concentrations in air of confined underground locations or in water of springs. A correlation between radon concentration and earthquakes could not been found so far [5]. One explanation to this is the very bad isolation of the measuring room. Outside temperature changes induce an important air exchange at the measuring place and thus the radon concentrations are lowered. Radon proves to be a very sensible tracer gas to tiny variations in the ambient, this explains the great interest of this gas. At night, when air exchange is minimal, radon exhalation from the ground can reach rather high concentrations in a very short time, showing a great underground radon potential. Thus these results showed that the measurement place is a most promising site for radon monitoring.

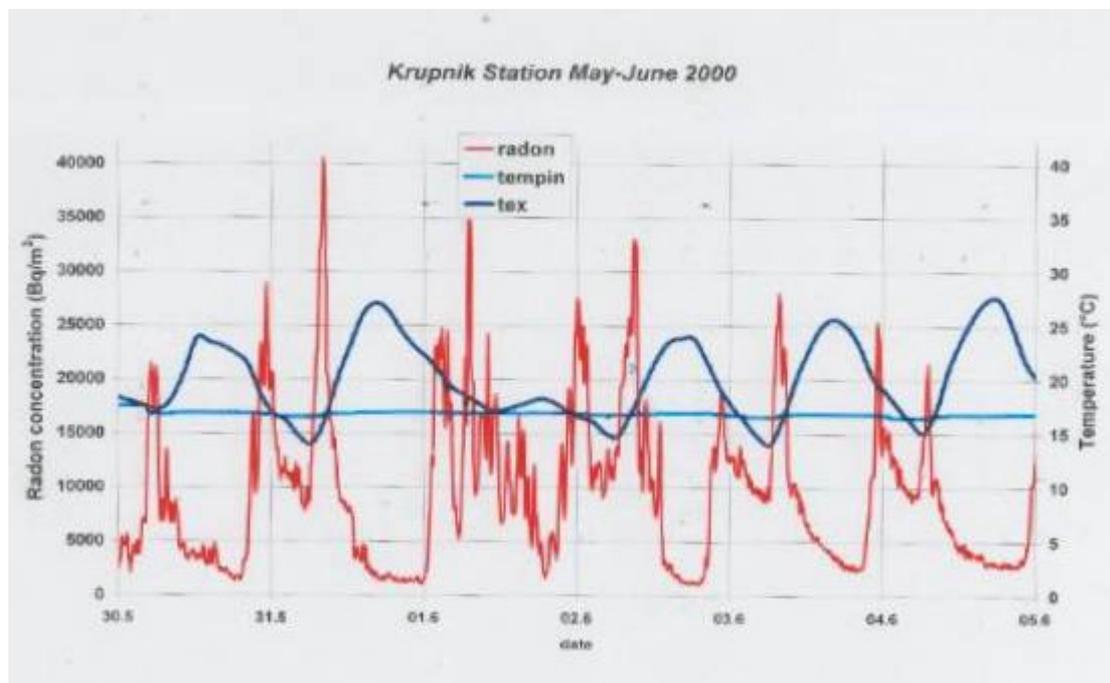

The air-soil samples around the station show radon concentrations ranging between 15 and 35 kBq/m3, these concentrations are much lower than the highest concentrations measured in or under the basement. An explanation for the high radon emission may be the fact that the station is located over a big fault zone [4]. It is known that usually over faults the radon concentrations in the air are increased due to the fact that radon are carried along with deep originated gasses. Due to the not isolated basements of the Kroupnik station, radon can migrate freely into the indoor air. As already noticed before, outdoor-indoor temperature differences (actually differences in air densities) and to a lesser extend air-pressure are essential to influence radon concentrations and may mask a possible influence of local earthquakes [6]. Radon as a tracer gas for the geodynamic investigations has been showed. This is a possible explanation for the high radon emissions in the site. The complex interpretation of the data from the radon measurements together with the other geophysical particularities show the intensive geodynamics of the Simitly graben and after scientific interpretation a local geodynamic model can be proposed [1,2,7].

The new measurements show clear daily variations due to the external temperature variations with a negative correlation.

Another fact has been recognised - the air pressure changes did not influence strongly the measured radon concentrations.

A very interesting behaviour of the radon concentrations has been observed in the confined place below the basement of the instrument room. A sharp and fast increase is followed by the sharp decrease after few hours. After that a stable phase occurs which lasts 3-4 times longer. If the increasing phase may be explained with the daily decrease of the temperature and the decrease phase by the diffusion and the dispersion of the radon in the outer space, the stable phase is more unusual. Something keeps the low emission stable. The possible explanation may be the accumulation of radon as a result of the increased outdoor temperature. Then a physical explanation model could be summarised. After the nightly temperature decrease, in the critical moment the radon transfer pathways are freed and the concentrated radon air is rapidly emitted in the surrounding space. Due to the not very nice closure of the confined space it dissipates in the space. Maybe, during the same time as a result of increased temperature the influx of radon charged air is inhibited and the emission is low, the concentration is low and the behaviour stable. After that phase the whole cycle starts again. The model looks simple but needs a further verification.

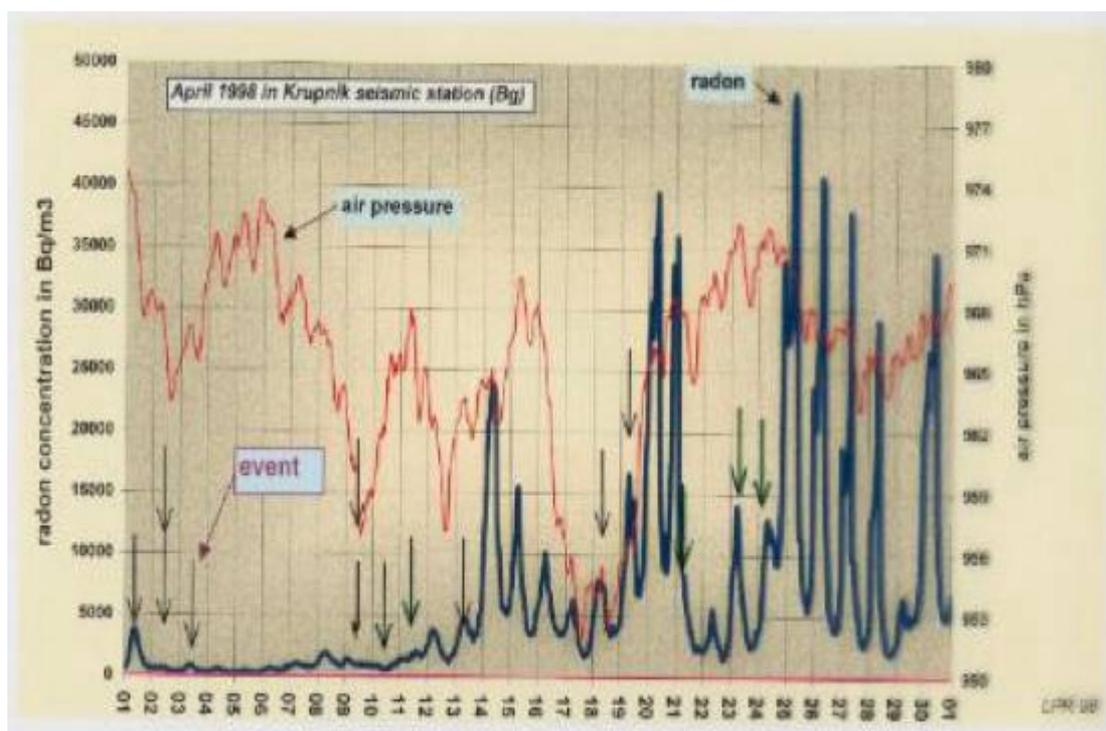

Recently a borehole with the diameter of about 8 cm and a depth of one meter has been drilled in the confined space described before. The upper 30-40 cm have been closed airtight, permitting only the passage of two tubes. By active pumping, each half-hour the air from the borehole interior is pumped through the tubing system into a closed chamber of an radon monitor. The preliminary results show fast and very big variations that seem to be able to confirm the suggested model.

The new Rn air measuring device with automatic registration is established now, and the new and more reliable data is under collection. The data will be proceeded in a near future, to check again the suggested model of Rn variations.

**Rn as a polluter and geodynamic indicator.**
With a half live of 3.8 days, radon is a short lived tracer gas. Radon could originate from the deep earth's interior. As an inert gas radon is very easily movable and in a fracture system can migrate and accumulate to high concentrations. This may explain the rather high concentrations measured in the basement of the Kroupnik station. Radon is a decay product of radium, it is the sole noble gas in the decay chain of uranium. Two short-lived decay products of radon are alpha emitters and, fixed on aerosols, are introduced to the respiratory tract. A high concentration of radon in the ambient air is supposed to be harmful, especially concerning lung cancer. There are recommendations for limiting radon concentrations in indoor air at homes and also in workplaces, 400 to 1000 Bq/m3 for homes. For workplaces the limit concentrations depend on the occupation time. As it was mentioned earlier, very high concentrations have been measured in the basement of the station. So this may be considered as a dangerous factor, if people were working there for a longer time. Our investigations show that in the upper flats where working and sleeping places are located,, the content of radon was lower than 300 Bq/m3 for the measured period.
An inset of air circulation in the basement shows that very rapidly radon concentrations are lowered.
So the general advice to the people working there are:
 - Not to stay and work for the long time in the basement.
 - Before entering there to make free air circulation by opening the door, windows and other air pathways.
 - It is good to make a free air circulation in the living rooms and working offices frequently as well.
These simple advice can protect fully the people from some possible negative influences of the radon as a radioactive polluter.
Easy effective measures can be performed for the people protection and isolation.

**General conclusions**
The Kroupnik seismic station can be considered as a good site for continuous Rn measurements for the geodynamic purposes. The results obtained previously show that for the geodynamic purposes the Rn is a good tracer gas and the station is a promising site for such research.
The people working and living there can be protected from the Radon polluted atmosphere in the basement of the station just following the simple rules for air circulation.
The continuos measurements in the soil, air and water is recommended to go on and to provide long-lasted sequences of the values for different interpretations and modelling connected with the local and regional geodynamics.

# МОЖЕ ЛИ СЪВРЕМЕННАТА НАУКА ДА ПРОГНОЗИРА ЗЕМЕТРЕСЕНИЯТА?

Ст.н.с. д-р Бойко Рангелов

Геофизичен инистут - БАН

Още от времето на възникването на сеизмологията (науката за земетресенията) в древна Гърция, хората са се опитвали да хванат едносричното - "да" или "не" относно възможността за прогнозиране на земетресенията, като винаги се предпочитали първото. За съжаление обаче и до ден днешен еднозначният отговор се изплъзва, както на учените занимаващи се с прогнозиранто на земетресенията, така и на всякакви псевдогадатели, самодейни "прогнозисти" и други представители на окултната или псевдонауката.

Известно е, че силните земетресения са бедствия с тежки последици върху хората и направеното от тях и в същото време се разразяват на неочаквани места и най-вече в съвсем непредсказуемо време. Още в древността, учените са забелязали, че след по-силно земетресение, следват множество по-слаби, но в същото време достатъчно силни трусове (наричат се афтершокове), които често предизвикват повече разрушения от основния трус. И са давали рецепти - след силно земетресение, да не се влиза в къщите и храмовете. А това, всеки ще се съгласи, вече носи елементи на прогноза.

Всъщност, сериозният научен подход, свързан с изследванията по прогнозиране на земетресенията, започва след II-та световна война. През 1949 година тежко земетресение разрушава град Ашхабад и Сталин нарежда да започнат задълбочени изследвания за прогнозиране на земетресенията. Така се слагат основите на Руската програма, която продължава повече от 45 години - до отделянето на средноазиятските републики, като самостоятелни държави. Веднага след това, в началото на 50-то години и в САЩ, на известния разлом Сан Андреас, започва разполагането на апаратура, която да помогне за решаването на основния въпрос - кога, къде и колко силно земетресение може да се случи в Калифорния. Днес, там е разположен най-големият и скъпоструващ комплекс от апаратура за прогнозиране на земетресения в света.

## КАКВО ПРЕЧИ НА ПРОГНОЗАТА?

Всъщност трите елемента на прогнозата ( кога, къде и с каква сила), представляват основния "препъни камък" на съвременната наука за Земята. И нейното най-голямо предизвикателство. Тук е мястото да се отбележи, че в резултат на дългогодишните изследвания на учените, днес е пределно ясно, къде и колко силно земетресение може да се очаква. Основният проблем е свързан с точното време на очакваното бедствие. И тук, нещата трябва да бъдат разделени. Съществуват различни прогнози - в зависимост от времето което покрият. Има дългосрочна прогноза (покрива време от няколко десетки години), средносрочна (от няколко десетки - до няколко години), краткосрочна (няколко години - до няколко месеца) и оперативна (от няколко дни или дори часове). Според съвременната наука, при достатъчно статистически данни, първите два вида прогнози са удачни във вероятностен смисъл. Така например, американските и руските учени ( по- голяма част от които са привърженици на хипотезата за "сеизмичен цикъл", за разлика от китайците например, които главно говорят за "сеизмичени епизоди") бяха изчислили в началото на 80-те години, че с вероятност 90% в Калифорния ще стане земетресение с магнитуд над 7.0 до края на века. Ако се погледне от тази гледна точка, земетресенията в Лома Приета (М= 7.1), и Нортридж (М=6.9), могат да се смятат за успешно предсказани. Това обаче с нищо не топли роднините на загиналите или собствениците на разрушенните и изгорели къщи.

Има и други случаи, когато бедствията се разразяват на места където не са били очаквани - примерите са много: Спитак (Армения 1987), Нефтегорск (Сахалин, 1995), Кобе (Япония 1995) и др. Хората искат, учените да им кажат, точен час, място и сила на очакваното бедствие. И по-възможност - по рано.

Е, точно тука, съвременната наука не може да им помогне, въпреки неимоверните усилия и средства които се отделят за това. А не може да го направи поради една много проста причина, наречена от учените

## "НЕЕДНОЗНАЧНОСТ НА ПРЕДВЕСТНИЦИТЕ"

Както вече беше казано, определени места с висока локална сеизмичност (наречени за краткост - полигони) се подлагат на детайлни геодинамични изследвания с най-съвременна апаратура - сеизмометри, лазерни далекомери, измерители на аномалиите на различни геофизични полета (магнитно, електрично, гравитационно и т.н), измервания на подземни флуиди (концентрация на радон или други газове с дълбочинен произход, темперетура, дебит на подземните води и др.) и други най-различни измервания, достигащи по някога да парадоксални наблюдения (над вид японски рибки - например, за които се смята, че излъчват електромагнитни импулси в зависимост от силата на приближаващото земетресение). Изобщо на съвременната наука са познати над 300 (а някои казват и 500) различни аномални явления, предхождащи силните земетресения - наречени предвестници и проявяващи се различно дълго време преди земетресението [1]. Е, точно тука е "заровено кучето". Оказва се, че нито един от тези предвестници, не се проявява при всички! наблюдавани земетресения, и обратното - съществуват аномалии в някои от известните предвестници, а земетресение - не става! А, защо това е така - днес никой специалист не може да каже.

## КАКВИ СА ФАКТИТЕ?

Примери за подобни ситуации има достатъчно. През януари 1976 година в Китай, в областта около гр. Хайченг започват де се наблюдават множество, известни преди това на учените, предвестници. Има и масови излизания на влечуги над снега, и издигане на нивото на водата в кладенци и помътняването й, и някои аномалии в земните елекрични токове и електропроводимостта на земята. Тези аномалии продължават повече от месец. Към началото на февруари, сеизмичните станции в района започват да регистрират повишен брой слаби трусове (учените често ги наричат форшокове - на български - предтрусове). По това време в Китай бушува културната революция, населението е военизирано и властите решават да организират евакуация на близо милион и половина души от града. Евакуацията се осъществява в продължение на три дни и на третата вечер, земетресение с магнитд около 7.3, изравнява със земята град Хайченг. Загиват едва 50-60 души - такива на които е омръзнало да студуват или не са се подчинили на властите. Това е първият на практика досега случай на прогнозирано земетресение с осъществена евакуация на населението. Само няколко месеца по-късно обаче, през юли край град Тангшан, подобно земетресение (М=7.8) [2] взема над 500 000 жертви. И за съжаление никой не е подозирал, че подобно бедствие приближава. Това кара специалистите да смятат, че успешната прогноза край Хайченг, е по-скоро плод на добро случайно стечение на обстоятелствата, отколкото научно издържана прогноза.

Тук не може да не се спомене и другият, станал вече учебникарски пример, за аномалията край Палмдейл [3]. През 1970 година, новоинсталираната апаратура в Калифорния край градчето със звучното име, показва много голяма по амплитуда деформационна аномалия. По изчисленията на специалистите, земетресението което може да се очаква трябва да е със сила над 7.0 по скалата на Рихтер. Наблюдават се и други аномалии - в отношението на скоростта на напречните и надлъжни сеизмични вълни, увеличава се броят на местните слаби земетресения, изобщо всичко говори, че наближава страшен катаклизъм. Учените са напрегнати, властите също не знаят да предприемат ли или не евакуация. Паниката без прекъсване продължава около 2 седмици. Към края на този период, едно нищо и никакво земетресение с магнитуд около 4.0 (подобни земетресения в Калифорния са ежедневие), става недалеч от наблюдаваните аномалии, а те постепенно затихват с времето [4].

Тези два най-фрапиращи примера са, за да покажат на неспециалиста значението на термина "нееднозначност на предвестниците".

Поради казаното по-горе, а съществуват и множество подобни примери, изброяването на които само би отнело място и време, учените понастоящем смятат, че съвременната наука не може да реши еднозначно задачата за кракосрочното и оперативно прогнозиране.

## МЕТОДЪТ ВАН

Напоследък твърде много нашумя т.н. метод ВАН (по името на неговите създатели - Варотсос, Алексополус и Номикос - гръцки учени от Университета в Атина). Същината на този метод се състои в измерване на земните електрични потенциали и в опита му да ги свърже по амплитуда и вид на аномалията, с времето и силата на приближаващото земетресение. В това разбира се няма нищо лошо - подобни измервания са правени в САЩ, Япония, бившия СССР и даже в България. Лошото идва от претенцията на авторите, че са открили универсален метод за прогнозиране на земетресения. Като доказателства за това се привеждат примери от изпратени телеграми до правителството на Гърция за точното време ( в рамките на едно денонощие), място и сила на очакваното земетресение.

## АНАЛИЗЪТ

на тези телеграми след публикуването им и съдържащата се в тях информация, проведен от гръцките сеизмолози обаче показва, че тези прогнози са в рамките на вероятностите и най-често не съответстват по място на очакваните земетресения. В цяла книга [6] посветена на този въпрос, ясно се вижда, че при огромното количество земетресения ставащи в Гърция, винаги може да се намерят такива, които "уж" съответстват на посочените. Тук е мястото да се каже, че двете най-силни земетресения, едното край Кожани (M=6,3) и в Патра (M=5,8), довели до жертви и множество разрушения, изобщо не фигурират в телеграмите на авторите на ВАН. Скандал избухна и когато беше публикувана прогноза за очаквано силно земетресение на територията на съседна на Гърция държава, която доведе до паника сред населението й.

## ПОЛОЖЕНИЕТО У НАС

След толкова обстойното запознаване със световната и европейска практика по прогнозирането на земетресенията, не е зле да се види и състоянието на този въпрос у нас. Още в началото на 80-те години започват методически проучвания. Появяват се спорадични изследвания и резултати свързани с вероятностния подход на средносрочното и краткосрочно прогнозиране. Изработен е вероятностен модел, който "работи" относително добре за силни ( с магнитуд над 7.0 ) земетресения в района на Егейско море [7]. Малко по-късно започват изследвания по търсене на предвестници - измервания на радон от източника на Софийската минерална баня, изследвания аналогични на метода ВАН (с някои специфични модификации) по измервания на земни електрични потенциали в сеизмична обсерватория "Витоша", както и измервания на концентрацията на хелий в района на Симитли в най-активната сеизмична зона на България - Крупнишката [8]. Там се разполагат и някои екстензометрични прибори за измерване на движенията по разломите на местната разломна система. Като резултат от тези опитно-методични работи са получени и някои интересни резултати. Например, преди първото Стражишко земетресение (месец февруари 1986) са наблюдавани положителни аномалии на земните потенциали в района на СО "Витоша" [9], корелационни зависимости между измененията в концентрацията на хелий и слабата сеизмичност в района на Крупнишката сеизмична зона   [10] и някои други. Имаше и проект за геодинамичен полигон София-Пловдив, но поради липса на средства, тази идея беше изоставена. Сега, отново липсата на средства е обрекла задачата за прогнозиране на земетресенията в България на застой [11].

## ИМА ЛИ АЛТЕРНАТИВА ПРОГНОЗИРАНЕТО?

В последно време обаче се появяват и идеи отричащи търсенето и разпознаването на предвестниците за краткосрочно прогнозиране на земетресенията [12]. Заговори се дори за края на експерименталните полигони, където ставаше това. Започва да доминира идеята, че земетресенията са свързани в единство с общите закономерности на катаклизмите ставащи по цялата планета. Предлага се създаването на обща мониторингова система, която използувайки постиженията на съвременната наука и техника, да обработва и анализира цялостната информация за ставащите природни бедствия по цялата Земя в реално време, със свръхмощни суперкомпютри свързани в световна мрежа. Основа за това е хипотезата, че всички аномални геофизични явления (каквито всъщност се явяват природните бедствия), са взаимно свързани и взаимно обусловени. С помощта на всеобщата система за наблюдение, със спътниковите средства за глобална комуникация и обмен на данни в реално време и със съответния анализ се предполага, че по-лесно ще бъде направено прогнозирането на очакваните природни катаклизми. Дали тази идея ще се окаже плодотворна, единствено времето и експериментът ще покажат.

# Earthquake prediction based on electrical signals recorded on ground surface.

**Dr. Thanassoulas, C.**
**Dept. of Geoph. Research,**
**Institute of Geology and Mineral Exploration (IGME),**
**Greece**

**e-mail :** thandin@otenet.gr
**URL :** http://users.otenet.gr/~thandin/index.htm

## Introduction

During the last one hundred years of instrumental seismology as far as it concerns the earthquake prediction topic little has been done. The main problem that was faced during all research efforts was mainly the nonlinear behavior of the earth, the chaotic conditions met at the seismically active area (fig.1)

## STRESS CHARGE MODELS

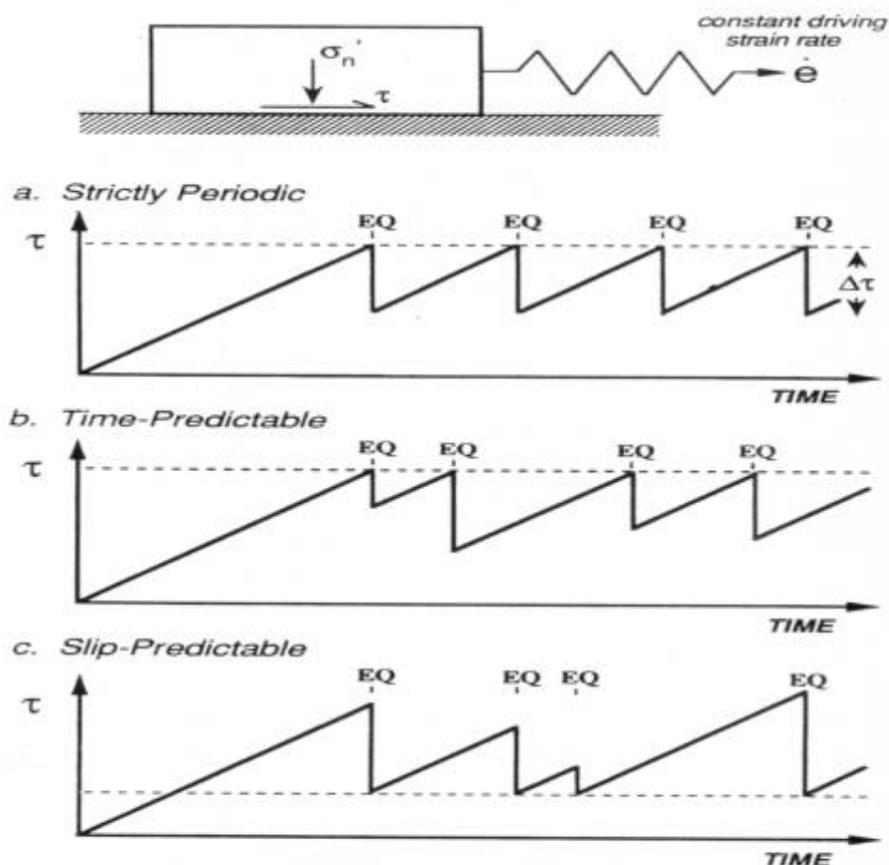

Fig. 1. Spring-loaded block-slider driven at constant loading strain rate, with three simple recurrence models for earthquakes (*EQ*) occurring by frictional stick-slip (after Shimazaki and Nakata, 1980).

Figure 1

and therefore the inability to postulate a generally accepted physical model that could account for the chaotic state of the focal area (Main, 1999). Moreover it has been demonstrated and presented in the relevant bibliography that earthquakes can occur almost everywhere and at any time so consequently these can be studied only with statistical methods. To this end several statistical methods have been developed and tested over wellknown seismically prone active areas but the results were only encouraging.

Aside to the topic of the earthquake prediction the long-term study of the occurrence of earthquakes over a wide regional area has been used quite successfully for identifying major tectonic features of the earth.

1. **Preliminary theoretical hypothesis for explaining the correlation to future local earthquakes**.

An earthquake occurs whenever the stress load of the seismically active area exceeds its fracturing stress level. The main mechanisms producing stress increase are a) the motion of the lithospheric plates that more or less causes a linear increase of it and b) the oscillation of the lithosphere due to earth tides (Knopoff 1964, Thanassoulas et al. 2001). This type of mechanism generates an oscillatory type of stress increase so that it is dependent to the oscillatory mode of the earth tides (12h, 24h, 14days, 6months, 1year periods). This has the effect that the critical stress load conditions required to trigger an earthquake are met mostly on the maximum amplitudes of the oscillation of the lithosphere, that is at the corresponding peaks of the earth tides (fig.2).

## TOTAL RESULTED STRESS – LOAD

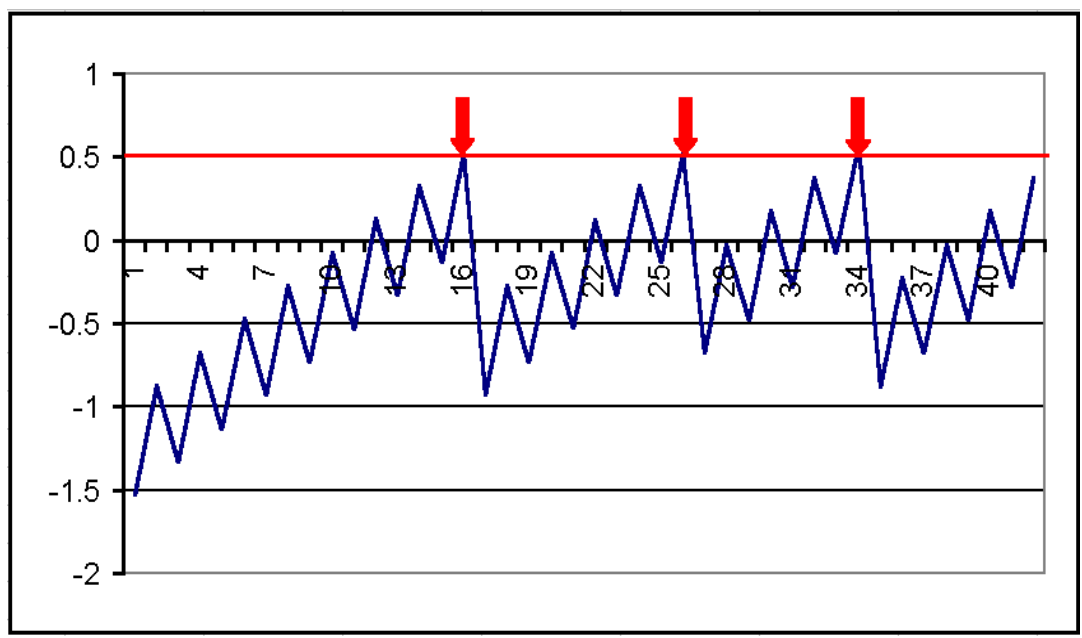

Figure 2.

Generally the electrical state of the earth is being described by a function U(Pi) that incorporates as main parameters Pi the position (X,Y) of the observation site, the time (t) of observation and any other parameter that can modify its form. This field is being modified before the occurrence of a strong earthquake due to the generation of electrical fields at the focal area, and consequently by measuring this difference it is possible to identify the presence of the "anomalous" field.

During the phases of extreme stress load various mechanisms are being activated that generate electrical signals. The induction of current at earth's surface (Meyer and Teisseyre, 1988) (fig.3).

# THEORETICAL MODEL
## a. Induction model

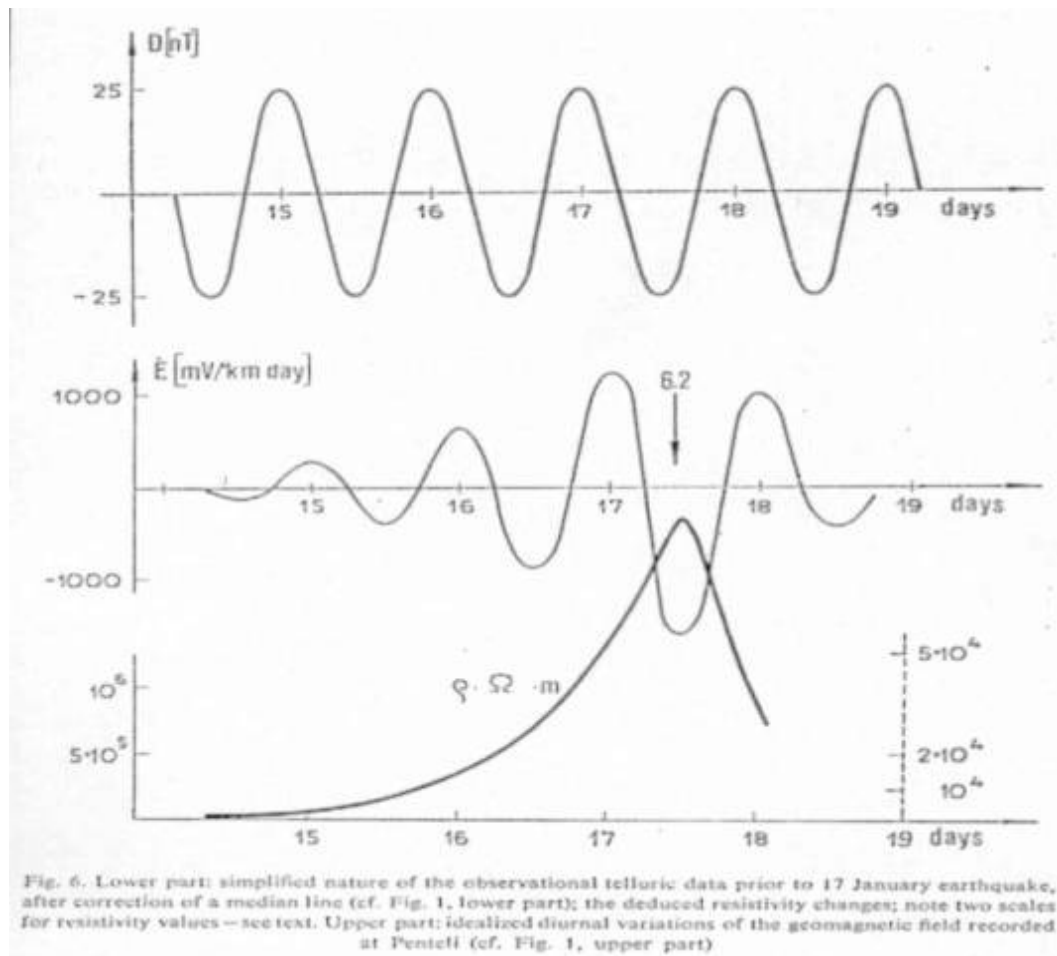

Fig. 6. Lower part: simplified nature of the observational telluric data prior to 17 January earthquake, after correction of a median line (cf. Fig. 1, lower part); the deduced resistivity changes; note two scales for resistivity values – see text. Upper part: idealized diurnal variations of the geomagnetic field recorded at Penteli (cf. Fig. 1, upper part)

Figure 3.

when combined to resistivity changes that precede an earthquake give rise to an oscillatory electrical signal that its amplitude increases continuously some days before the occurrence of an earthquake. After the occurrence of the earthquake its amplitude decreases drastically. Another mechanism is the activation of the piezoelectric properties of the lithosphere (Thanassoulas and Tselentis 1993, Client 1999) due to is large content of quartzite. This mechanism gives rise to various electrical signals (fig.4)

# THEORETICAL MODEL
## b. Piezoelectric model

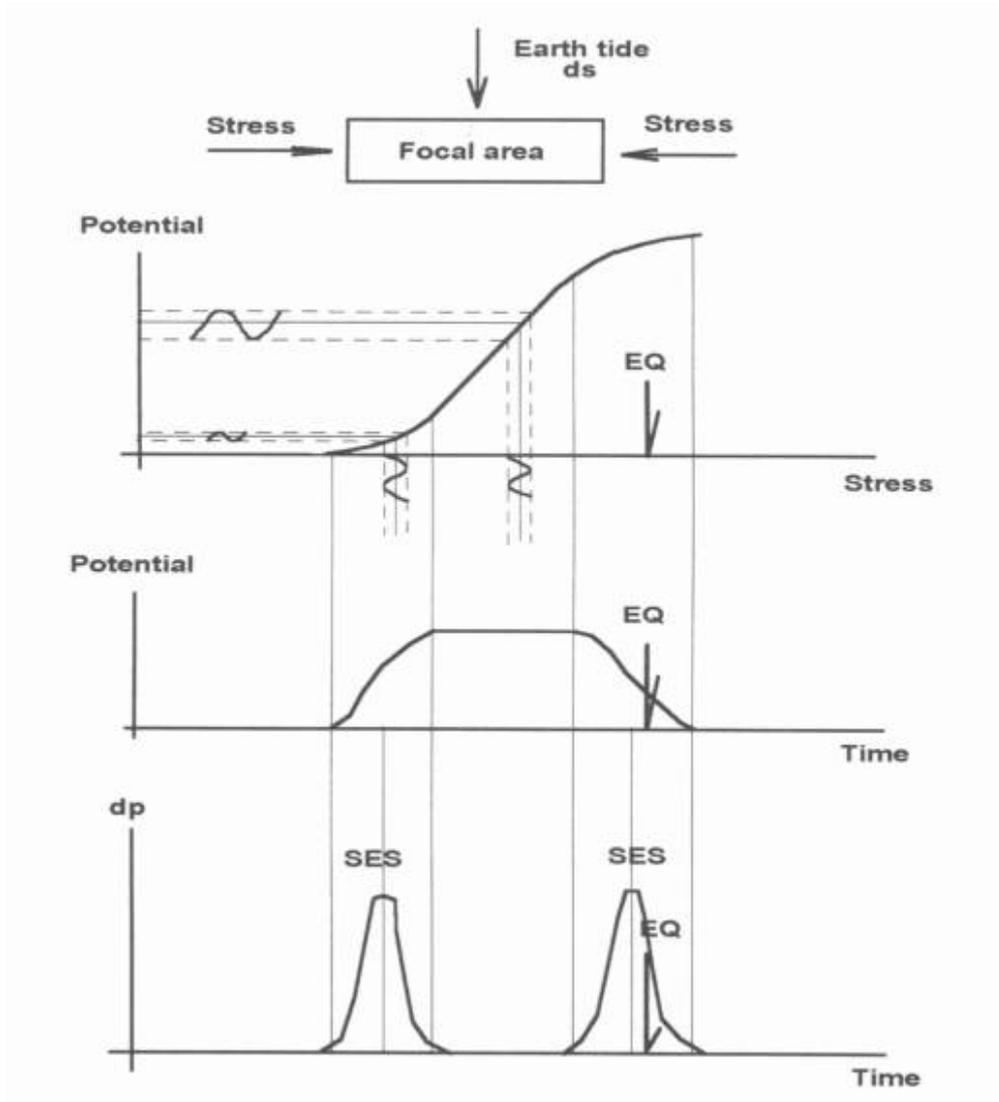

Figure 4.

An oscillatory one with 24h period is being generated due to the earth tides acting on the dynamic equilibrium-state of the focal area. Moreover due to the specific characteristics of the of the stress increase - piezoelectric potential function, a very long period (VLP) signal is being generated closely dependent on the first derivative in time of this function. Shorter period signals are being generated from second order derivatives of the same function. It is still well-known that rock fracturing generates electrical signals (fig.5)

## THEORETICAL MODEL
### c.  Rock fracturing model EM generation

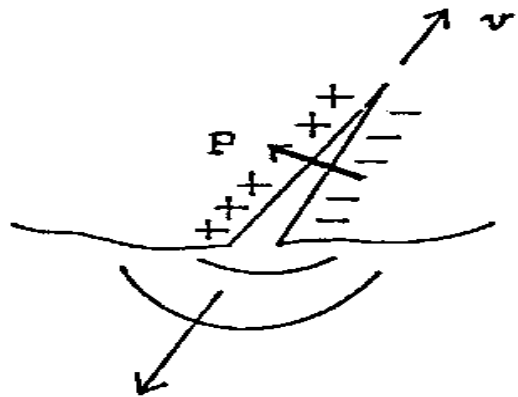

A model of rock fracture radiating EM waves.

Figure 5.

due to the fact that electrical charges at the edges of fracturing create short period current pulses. VAN group has presented a mechanism (Varotsos et al. 1984a,b 1990) that generates piezostimulated currents (fig.6)

## THEORETICAL MODEL
### d. Piezostimulated currents

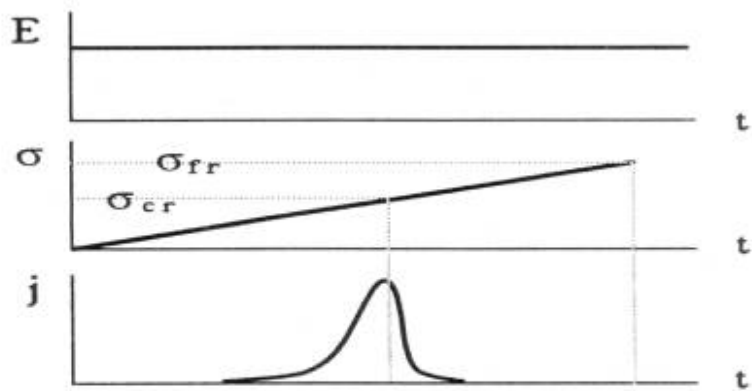

The piezo-stimulated current, $j$, flows during a stress accumulation stage at a critical stress level $\sigma_{cr}$, which is smaller than the fracture stress of rocks $\sigma_{fr}$, while the external electric field, $E$, is kept constant (after Varotsos and Alexopoulos, 1986).

Figure 6

under certain stress load conditions of a rock sample, well before its fracturing takes place.

## 2. Apparatus and measurements methodology

The hardware used to implement this method at Volos (Greece) monitoring site (Thanassoulas and Tsatsaragos, 2000) consists of two receiving dipoles (Fig. 7) at NS and EW direction, and of almost 100 meters length each.

### MONITORING SITE AT VOLOS AREA – CENTRAL GREECE

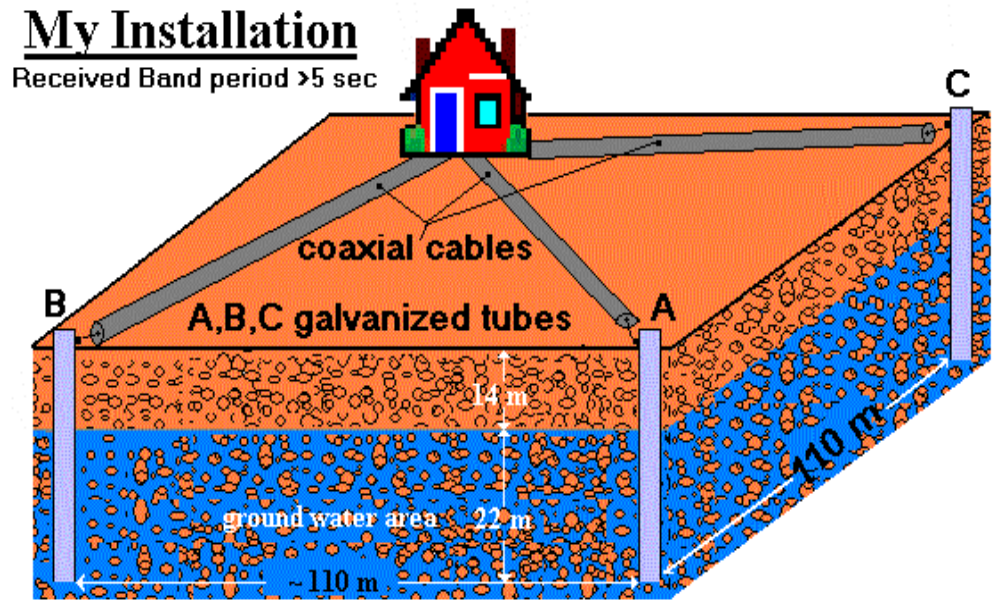

Figure 7

The analog signal is split in two parts. The first one is low pass filtered at a cutoff period of 5 sec and the second one is band pass filtered at a center frequency of 24h period and cutoff of 12h. These signals for both directions are being sampled at a minute period and converted into digital form and stored in the memory of a computer.

These recordings are being processed by various techniques so that the anomalous signal is being separated. The azimuthal calculation of the electrical field intensity vector is being calculated for all the signal length and its azimuthal distribution is being performed and presented on an azimuthal circle. The average value of all these azimuths calculated represents the true predicted azimuthal direction of the earthquake to occur.

## 3. Results

This method has been a posteriori applied (Thanassoulas 1990, Ifantis et al. 1993) so far on data provided by VAN group on 1990 and 1993. In the first case (fig.8).

## AIGHION (Ms=6R) / KALAMATA (Ms=6.2R)EQs

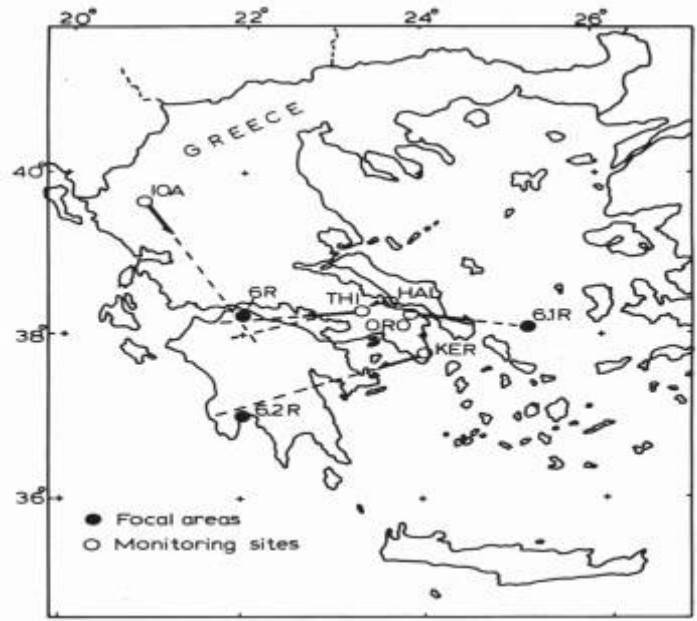

Figure 8

 three large Eqs were used from which the one of Eghion (Greece) was located using three electrical vectors while for the two remaining only the azimuthal direction was estimated due to the fact that only one monitoring site was available. In the second case (fig.9a,b)

## VOLOS EQ (Ms>6R)

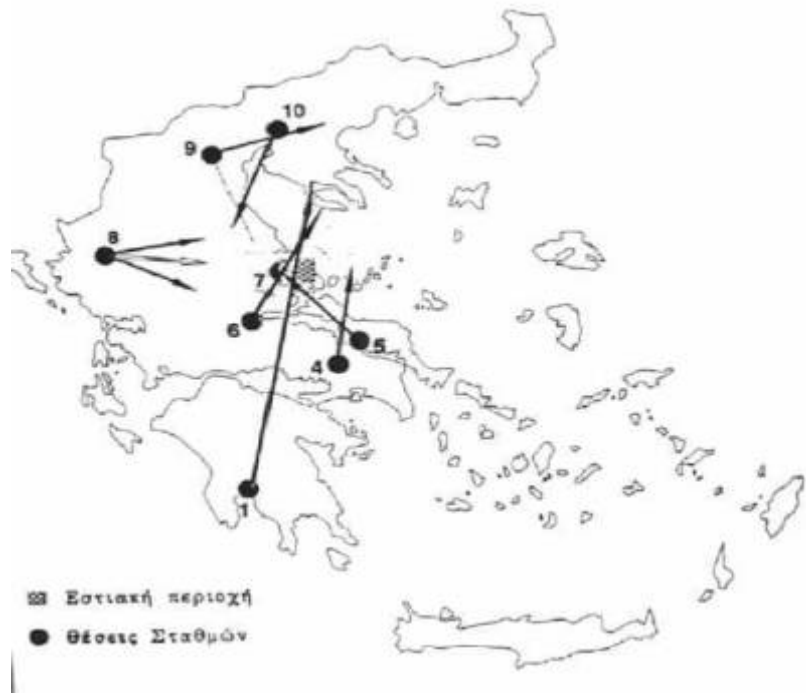

Figure 9a

## VARTHOLOMIO EQ (Ms>6R)

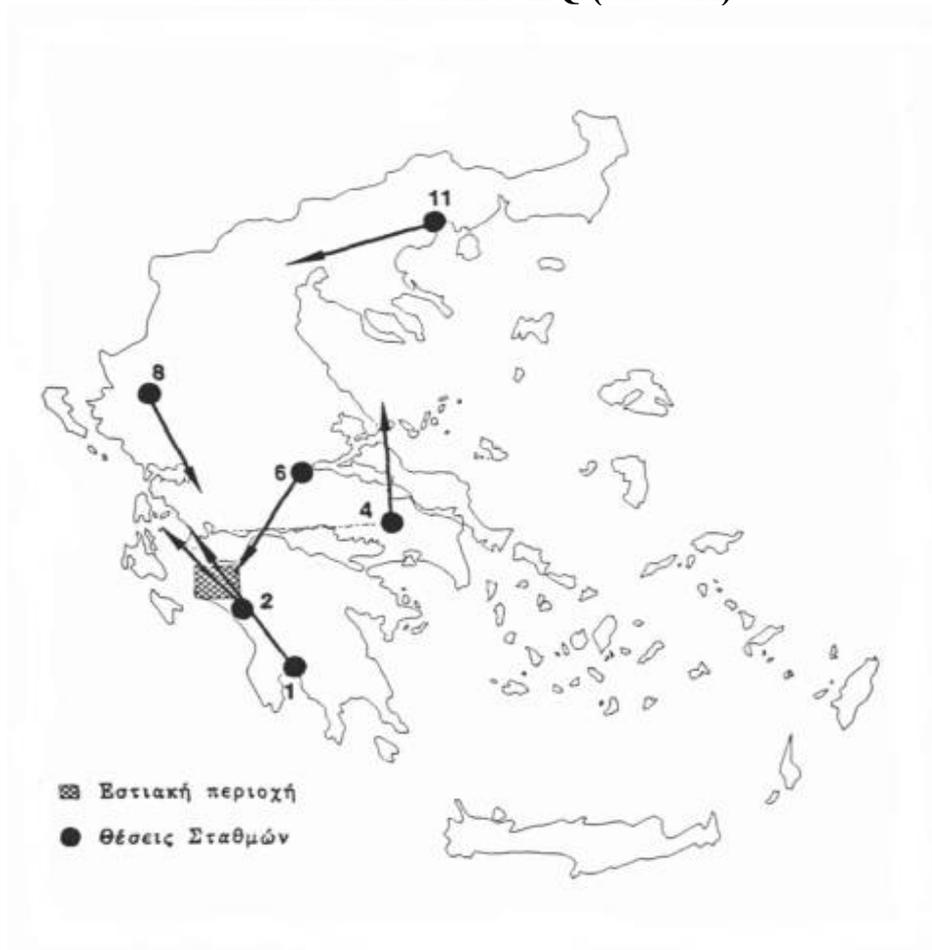

Figure 9b

two more large Eqs had been located by triangulating more than three electrical vectors from corresponding monitoring sites. The calculated probability for a by chance success is being calculated as 1/8000 for the case of three converging electrical vectors, while this probability decreases to 1/160000 in the case of four converging electrical vectors. Both cases for 1990 and 1993 epicentral area determinations, because of the extremely low value of the by chance probability calculated indicate the validity of the method.

The same methodology was applied on real data obtained from Volos monitoring site. In this case only the azimuthal direction of the imminent earthquake has been calculated since only one electrical vector is available. The following figures (fig.10a,b,c)

# RHODOS EQ (5.6R) – VLP SIGNAL

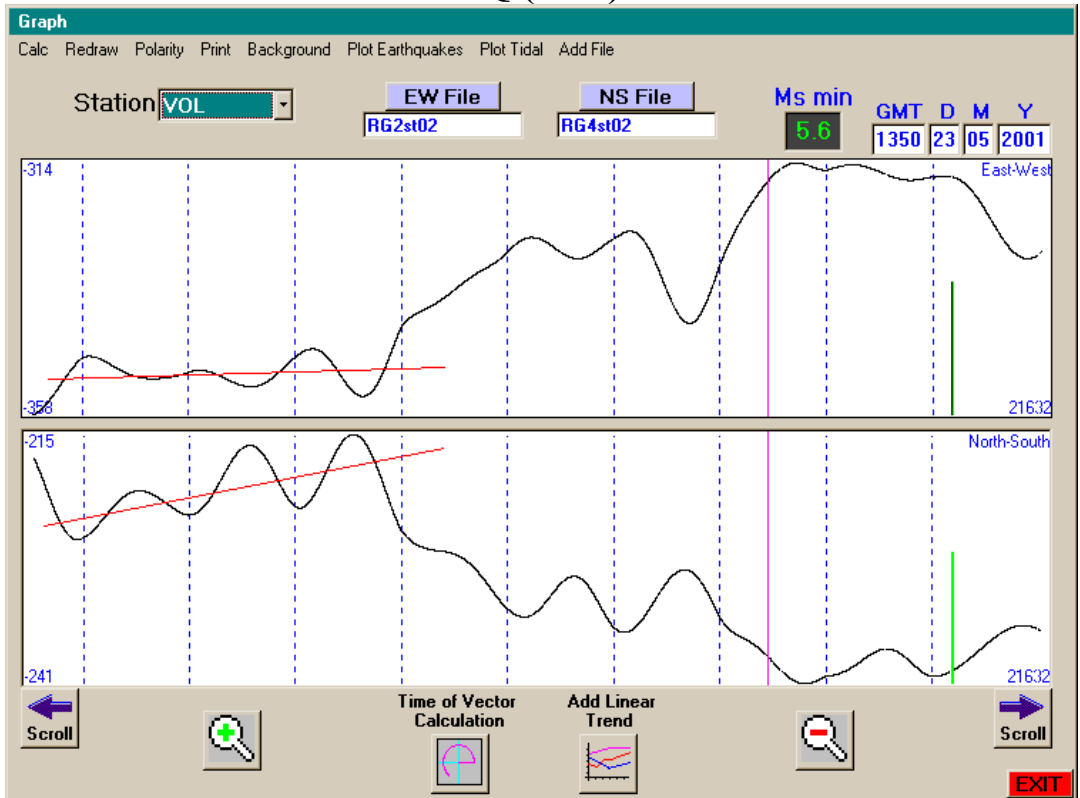

Figure 10a

# RHODOS EQ (5.6R) POLAR DIAGRAM

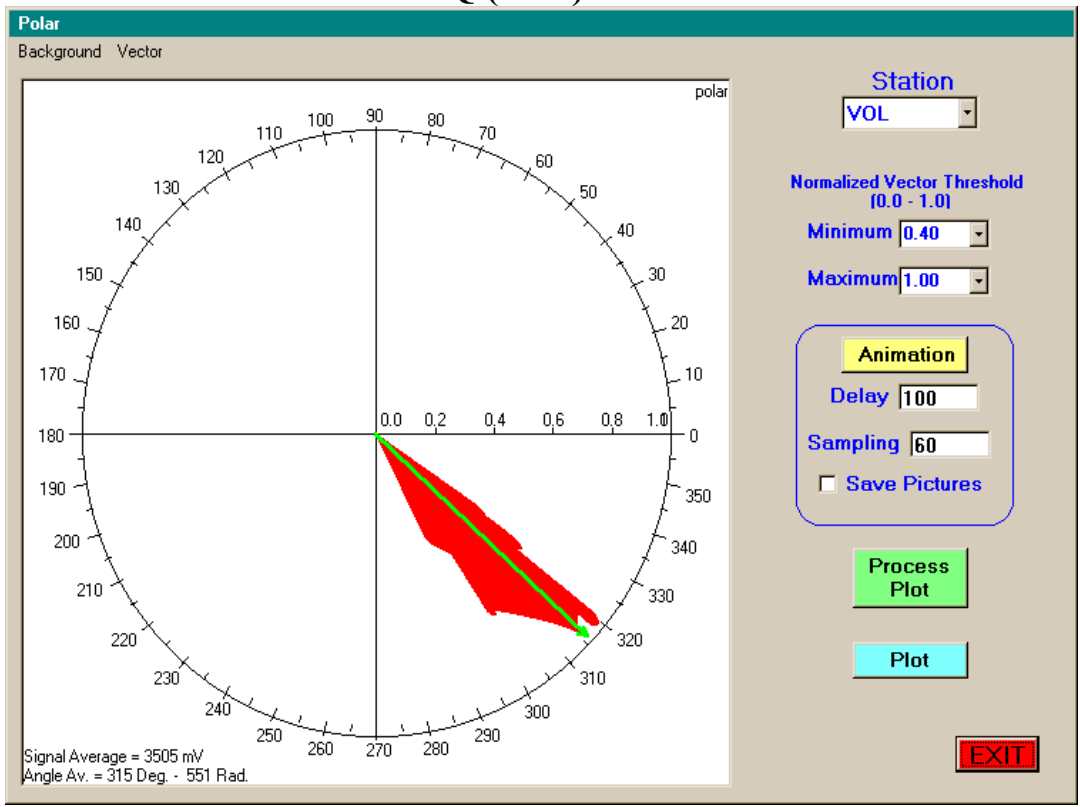

Figure 10b

**RHODOS EQ (5.6R)**
**AZIMUTHAL DIRECTION OF ELECTRICAL FIELD**

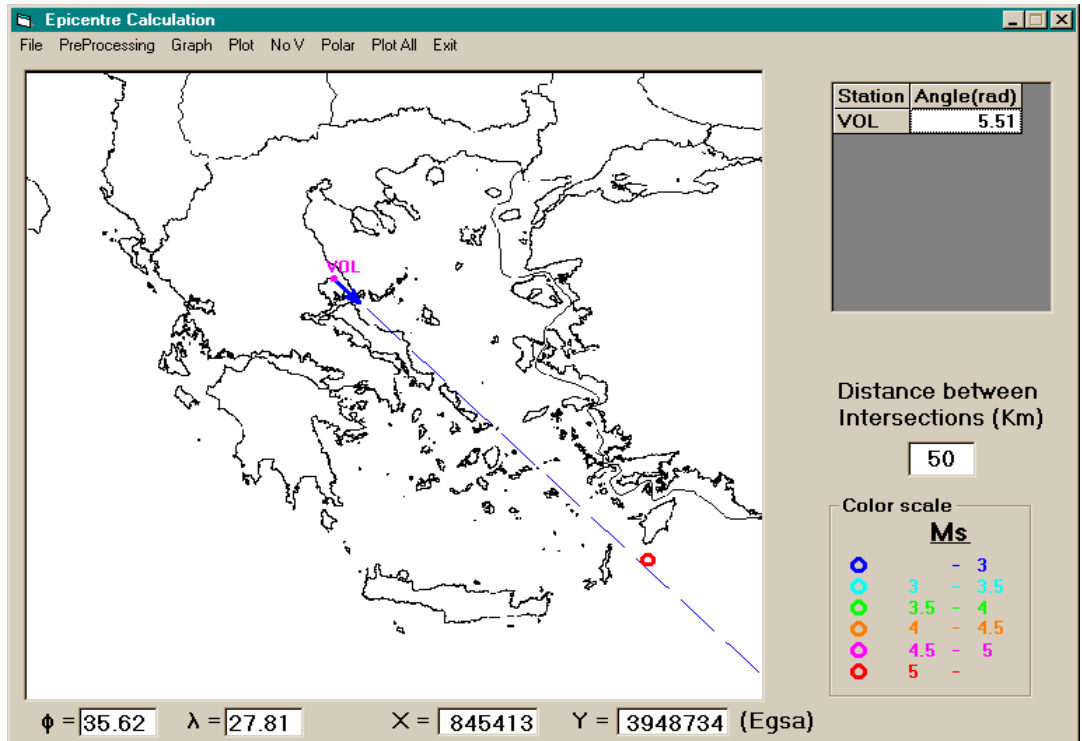

Figure 10c

corresponding to VLP signal, polar distribution of the electrical vector present the results obtained by processing the actual recordings compaired to the true location of the corresponding earthquake as it has been determined by the use of seismological methods from the Geodynamic Observatory of Athens Greece.

Moreover the time window of the occurrence of the imminent EQ is being estimated by studying the earth tide oscillation and determining its daily - monthly successive maxima and minima (fig.11a,b)

**DAY WINDOW DETERMINATION**

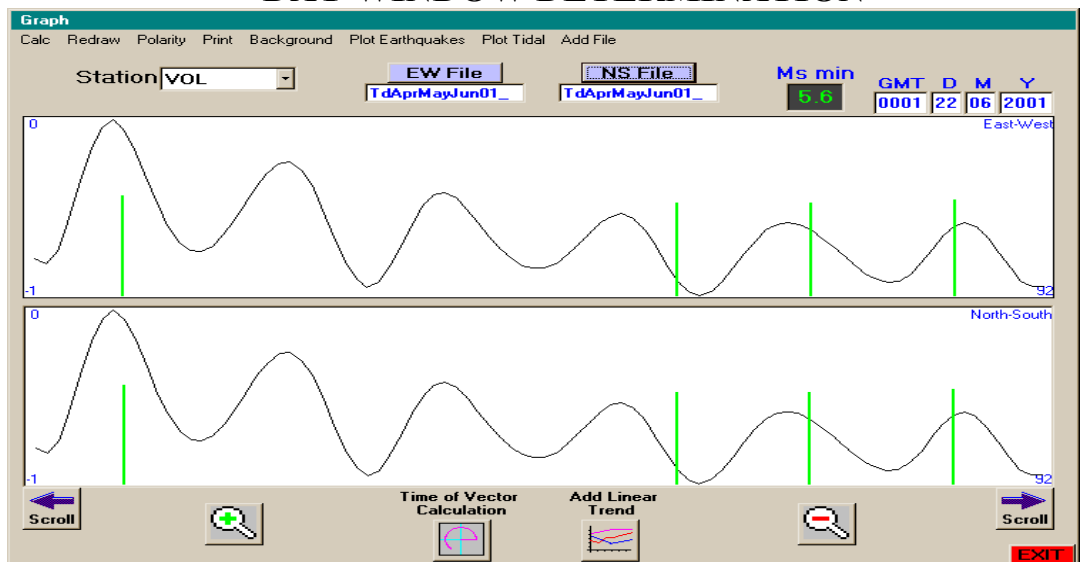

Figure 11a

## KONITSA EQ (5.8R)

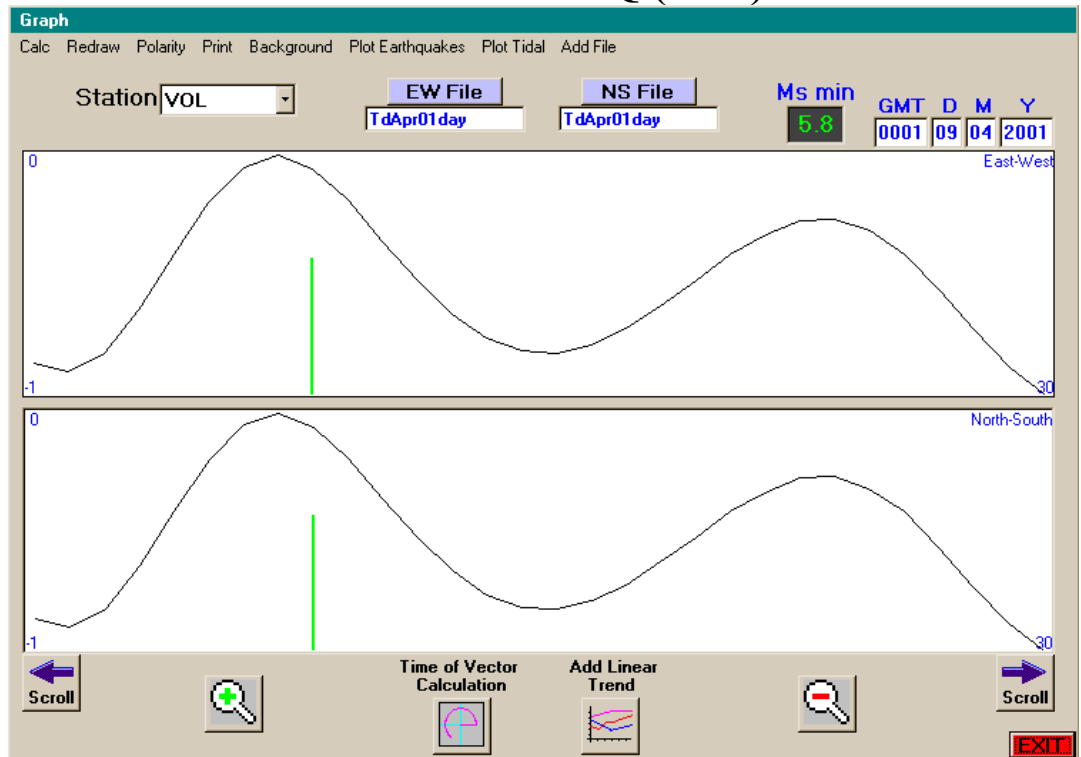

Figure 11b

that are precisely defined a priori. According to the theoretical models presented earlier concerning the increase of the stress load of the seismically active area consequently it is expected that these predefined tidal peak times are the most possible times that the imminent earthquake will occur. Finally the magnitude of the imminent earthquake has to be estimated. This is difficult at the present time since it is required a rather large regional network to be deployed and for a long period data to be collected for calibrating the entire network parameters to be used for the magnitude estimation.

### 4. Conclusions

The study of the results, obtained from earlier a posteriori application of the method on VAN data and more over from recent experiments on data obtained from Volos Greece monitoring site, clearly indicate that it is possible to calculate the epicentral area of an imminent earthquake by identifying the anomalous electrical signal generated at the focal area and by applying triangulation techniques on the corresponding electrical field intensity vectors of all monitoring sites.

The evolvement of electrical signals at a certain time indicates that the last phase of severe deformation of the crystal l lattice of the rock has been reached and this indicates intrinsically that the time window has been set to a few days. Further detailed study of the daily, monthly and sometimes yearly state of oscillation of the earth tides provides very accurate information of the most probable time of occurrence of the imminent earthquake.

### 5. Proposal for regional Balkan Network

For the implementation of a fully functional network it is clear that some more, at least 2-3, monitoring sites must be installed widely spaced over the area to be covered. The latest years an increased seismic activity has been observed especially in Turkey and Greece and generally over the Balkan Peninsula. Consequently it is considered of most importance for this network to be implemented as soon as possible. The participation of the Balkan countries will facilitate its installation and will foster the scientific cooperation (on the topic) of the scientific potential of each one of them.

**The possible correlation's between electromagnetic earth surface fields and future local earthquakes**

Author: Ass. Prof., Dr. Strachimir Cht. Mavrodiev
Institute: Institute for Nuclear Research and Nuclear Energy, Bulg. Acad. of Science

1. The History remarks

In the late 1970-th, there was the first precedent in known history of our Civilization, when the World leaders followed the Science recommendation of "Nuclear winter model" to start a new Era without nuclear, chemical, biological, geological and hydro-meteorological weapons. Now (2001), one can say the "Nuclear winter model" was the first example of the new "Science for sustainable development (Harmonic existence)"[1].

At this time started the formation of International science group from the Black Sea countries, connected with the Joint Institute for Nuclear Research, Dubna, which aim was to formulate the Science recommendation for preventing the "future" then ecological catastrophe of the Black sea ecosystem. In 1991 this work was done more or less [2].

2. Introduction

The title of this paper is a part of "Dynamics and correlations of proccesses and interactions in the Black sea ecosystem in real time" based on the complex Monitoring of phisical (including hydro-metheorological and geological), chemical and biological parameters of the Invironmental.

The concrete idea for looking the correlations between the earthquake and elektromagnetic fields was the attempt to explain the fire column in the sea observed at time of Crime earthquake. Our hypothesis was that the fire column was sediment methan hydrates, dehydrated because of earthquake pressure wave and fired from the high ionization (enough high temperature) of air column. From common point of wiev it seems natural the electric parameters of earth peel to be changed before and in the time of earthquake.

So, the accuracy measurement of the earth and atmosphere electrical curents and potential and the geomagnetic field in apropriate space and time scale, the analisys of the data should have a visible and measured correlations with earthquake processes.

The Programs for Erthquakes prediction, started after the American, China and Armenian grate earthquakes in 1980-th was closed. My opinion for this unsuccess is that The Programs [3] dealed only with geological parameters. My believing is that including of the elektromagnetics fields parameters can give a new live of the earthquake prediction investigations.

In the end of 1999 in the INRNE started the measuring of one geomagnetic field projection (one sensor) with accuracy less than 1 nT and at every 0.25 s. The sensor and elektronics device are "Know-How" of JINR, Dubna (Boris Vasiliev). The calibration was tested in Geomagnetic Station near Koprivchtica town, Geophysical Institite, Bulg. Acad. of Science. The solving of nonlinear problem for comparison of our dimension magnetic field behaviour with the standart vector measured in Koprivchtica, gave the Fourrier analysis time 1 minute.

The two-years monitoring shows the one can see all biggest world earthquakes (bigger than 8R) and the big and near enough local (Vranca, Greace, Macedonia, Turkey.) earthquakes.

3.  The results.

The evidence for correlation of MF behavior and the time of the earthquake occurrence has been classified in two categories.

The first one contains seismic events, in which the earthquake magnitude is bigger than 8 Richter. The behavior of MF is going to be flat for some time close and after the earthquake occurrence. From six such cases all over the World from 2001 to date four of them indicated clearly this distinct behavior, one with doubt and one without signal.

Table 1

| Year, | Month, | Day, Origin time, | Lat, | Long., | Depth, | Mag (R) |
|---|---|---|---|---|---|---|
| 2000 | 06/04 | 16:28:26 | -4.72 | 102.09 | 33 km | 8.3 (?) |
| 2000 | 06/19 | 14:44:13 | -14.8 | 97.44 | 10 | 8.0(No data) |
| 2000 | 11/16 | 04:54:56 | -3.98 | 152.17 | 33 | 8.2 |
| 2000 | 11/17 | 21:01:56 | -5.5 | 151.79 | 33 | 8.0 |
| 2001 | 01/26 | 03:16:14 | 23.42 | 70.23 | 16 | 8.0 |
| 2201 | 06/23 | 20:33:14 | -16.22 | -73.60 | 33 | 8.4 |

The second are for regional Balkan seismic events with Magnitude above five Richter (5R). From 16 such events, 9 of them presented similar behavior as in the first case, including some variations of oscillating character with hours period. Seven of them have no correlation.

Table 2.

| Year, | Month, | Day, Origin time, | Lat, | Long., | Depth, | Mag (R) |
|---|---|---|---|---|---|---|
| 2000 | 04/06 | 00:10:38 | 44.74 | 26.58 | 132 km | 5.4 |
| 2000 | 04/21 | 12:23:10 | 37.84 | 29.33 | 33 | 5.5 |
| 2000 | 05/24 | 05:40:37 | 36.04 | 42.01 | 33 | 5.9 |
| 2000 | 05/24 | 10:01:44 | 35.82 | 42.10 | 33 | 5.2 |
| 2000 | 06/13 | 01:43:14 | 35.15 | 47.12 | 10 | 5.4 (?) |
| 2001 | 05/01 | 06:00:56 | 35.17 | 47.50 | 33 | 5.1 (?) |
| 2001 | 05/24 | 17:34:01 | 45.69 | 26.42 | 141 | 5.3 (no) |
| 2001 | 05/29 | 04:43:57 | 35.41 | 27.78 | 20 | 5.1 (no data) |
| 2001 | 06/10 | 13:11:04 | 38.59 | 25.63 | 33 | 5.3 |
| 2001 | 06/23 | 06:52:41 | 35.71 | 29.01 | 33 | 5.3 |

Furthermore, two more seismic events located near the Sofia monitoring site presented the same behavior, despite their small magnitude:

Sofia,     2001/04/03     18:20, 2,3 R ;
Vranca,    2001/04/09     04:29, 395N 66,89E,  Dept 11.4km   Mag. 3.2 R.

The figures illustrates the usual and earthquake day geomagnetic field projection behavior:

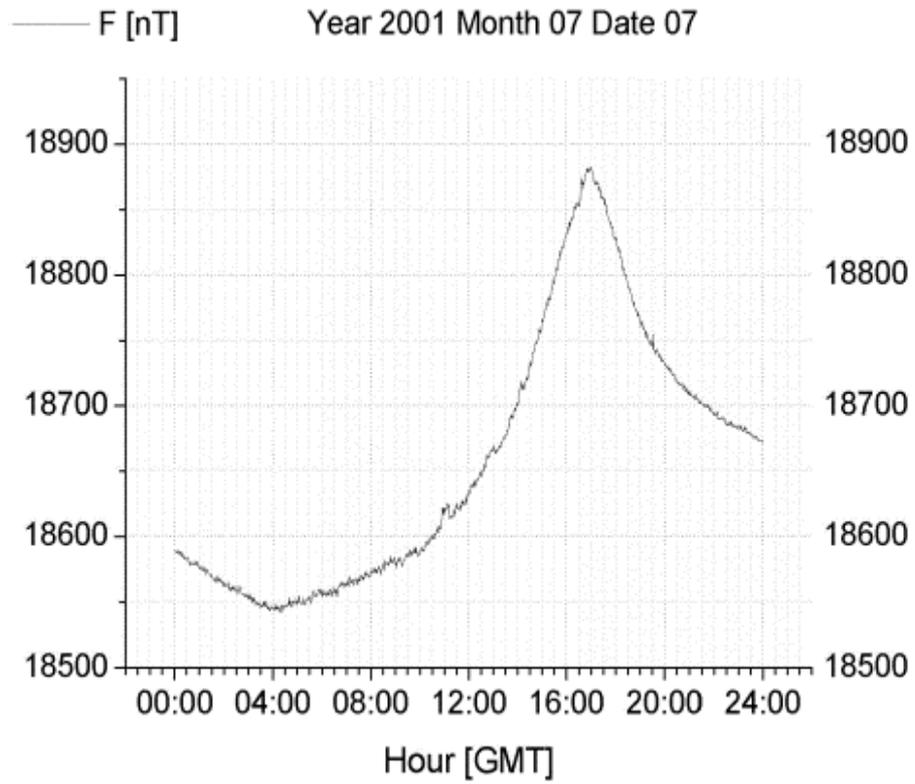

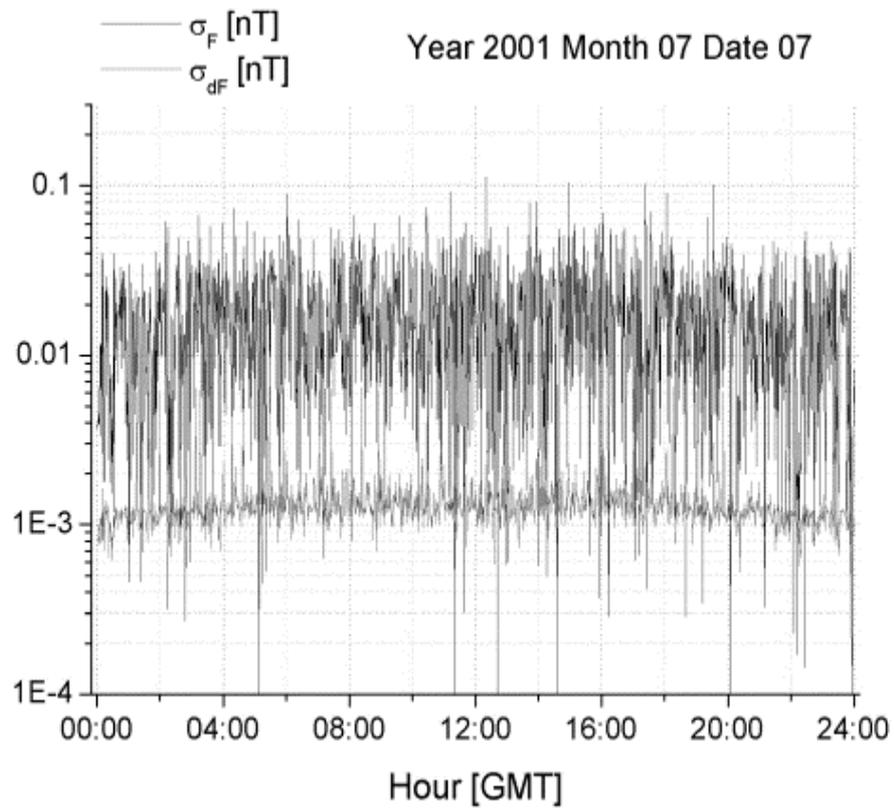

Fig.1 Normal day behavior

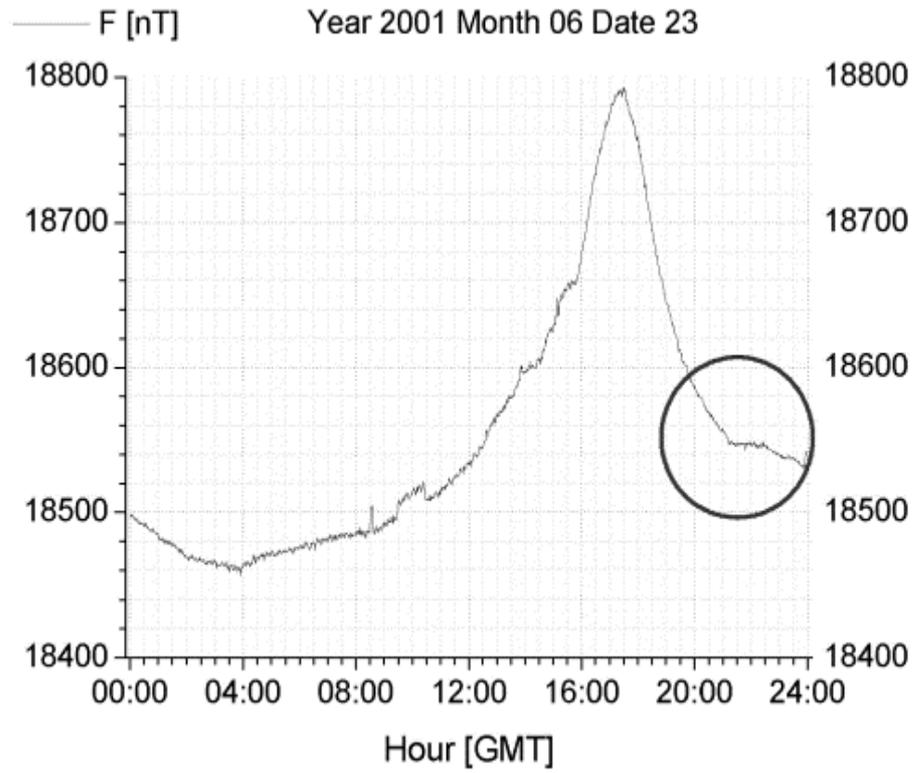

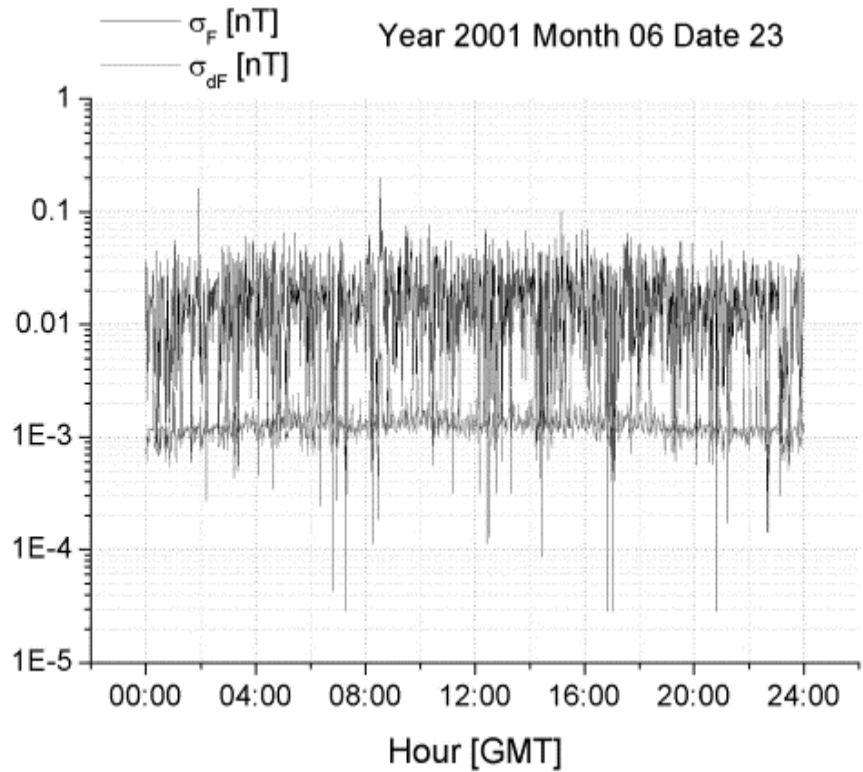

Fig.2 The flat behavior (in red circle) after the event

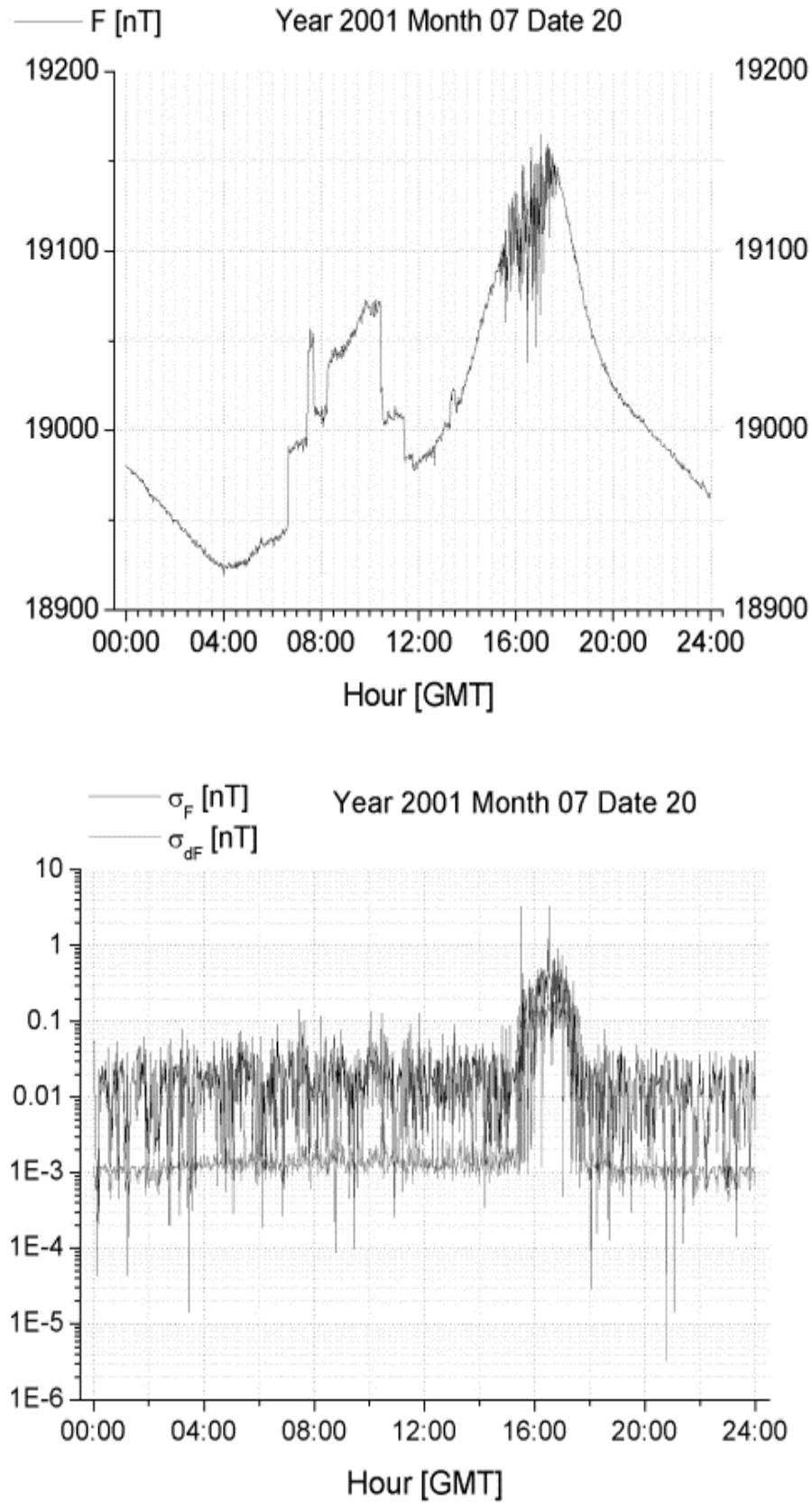

Fig. 3. The hours and minutes oscillations caused by a local event

4. Propose for monitoring

The new monitoring sites have to include:

The two components measurement (EE2) of earth electric potential and current (Thanassoulas current site);

The vector (tree components) Geomagnetic field measurment (GMF3);

One dimensional atmospheric electric potential and current measurement (AE1).

When the future epicenter aria is detected from EE2, one has to start in this aria GMF3, AE1 and other possible geological, geo-chemical and geo-physical parameters monitoring. The preliminary results permit us to believe it will be possible to measure the correlation's between the EE2, GMF3, AE1, daily solar wind influence and local earthquake parameters (time, power and depth) in obviously manner.

For the first part of the Investigation, for my opinion, the best aria is the Augean Sea, where we need to probe four Sites around the Sea (Volos,  )

If in some time (2-3 years) the predicted events will start to take place with probability near to 1, the Balkan region Earthquakes Prediction Monitoring will need at last 20 Sites.

Nevertheless, now one can say that we have to do The Balkan gravity 3 to 3 map.

The preliminary estimation cost of one Site is around 20000 US$, the time-around 6 month. The time for the first part of the Project (Pilot part) is around 1 year. The total cost, including 3 sites and Laboratory for archiving and analyzing the data, will be around 100000 US $.

At the end I would like to stress, that the Proposed project have some chance for success only in the case of united Balkans scientists, governments and local authorities.

## Seminar Conclusions

On the 23$^{rd}$ of July, 2001 a 3-day seminar was held in INRNE, Bulgarian Academy of Sciences, Sofia, Bulgaria, titled:

**"The possible correlation between electromagnetic earth surface fields and future earthquakes."**

During this seminar, **Assoc. Prof. Ranguelov, B., seismologist, Assoc. Prof. Mavrodiev, S. Theor. Phys. and Dr. Thanassoulas, C., Geophysicist** presented their research results on the seminar topic.

Furthermore the aim of the seminar was to investigate the possibilities for submission of a common research program, to establish a regional network for monitoring different geophysical parameters.

The presented examples of measurements of the earth's electric and geomagnetic fields indicate that it is possible to organize such type of a regional Balkan monitoring network for physical and geophysical fields that is going to permit the improvement of the earthquake prediction.

This was validated in practice by the occurrence, during this seminar, in Greece of a large earthquake (Ms=6.1R, 25/07/01) as it had been stated during the presentation of Dr. Thanassoulas and following the corresponding theoretical part of it (time determination), that could happen with very large probability. The magnetic observations, a few days before this large earthquake, presented unusual behavior too, with the specific daily variation of the 25$^{th}$ of July.

To this end the scientific community in Balkan region will be asked to collaborate into the implementation and submission of this project for funding by any appropriate authorities.

For easing the utilization of all the afore mentioned ideas the Organizing Committee proposes the foundation of a non government organization (NGO) under the title:

**"The Balkans, Black Sea Region Sustainable Development (Harmonic Existence) and Science."**

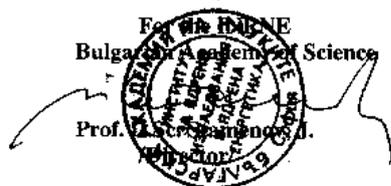

For the INRNE
Bulgarian Academy of Science

Prof. Apostolov, A.

The Organizing Committee

Assoc. Prof. Mavrodiev, S.Cht. (Bulgaria)

Dr. Thanassoulas, Geophysicist (Greece)

27 July, 2001
Sofia, Bulgaria

# PROJECT DESCRIPTION (1<sup>st</sup> draft)

1. **Bibliographical study of the status of the topic of the earthquake prediction - data bank organization.**

2. **Monitored parameters decision**

3. **Monitoring sites selection**

4. **Monitoring network utilization**

5. **Results dissemination**

1. **Computer library  : hardware + internet**

2. -<u>**Gravity unified map**</u> **(grid 3x3 or less)** for the contemporary mapping of the lithospheric deep fracture zones under the monitored regional area.

   - **Earth's electrical field variations** for the utilization of the triangulation of the electrical signals caused by large imminent earthquakes.

   - **Earth's- atmospheric electrical vertical field variations** over the detected future epicentral areas by triangulation

   - **Geomagnetic field measurements** for studying its correlation with EQ events and the way these Eqs modify the geomagnetic field.

   - **Noise sources study** that interfere the monitored geophysical fields (mainly the influence of solar wind)

   - **Resistivity distribution** in the lithospheric plate.

   - **Physical model built up** for the establishment of the correlation between the different physical parameters.

   - **Theoretical estimation of the expected anomalous values** of the monitored physical parameters based on the postulated physical model.

List of participants

From Greece

Institute of Geology and Mineral Exploration, Athens
Dr. C.Thanassoulas

From Bulgaria

Institute of Geology, Sofia, Bulgarian Academy of Sciences
Prof. DrS. D.Evstatiev
Assoc.Prof. DrS. S.Shanov
Assoc.Prof.,Dr. B.Rangelov

Institute of Sollid State Physics, Sofia, Bulgarian Academy of Sciences
Prof.Dr.S. D.Uzunov

Institute of Nuclear Research and Nuclear Energy, Sofia, Bulgarian
Academy of Sciences, www.inrne.bas.bg
Assoc.Prof., Dr. S.Donev
Eng. I.Kalapov
Eng. Ch.Lenev
Assoc.Prof.,Dr. S.Cht.Mavrodiev
Mgr. A. Mishev
Bac. L.Staikov
Prof. DrS. J.Stamenov
Assoc.Prof.,Dr. D.Trifonov
Dr. B.Vachev
Assoc.Prof.,Dr. S.Ushev